\documentclass[twocolumn, pra]{revtex4}
\usepackage{amsmath, amsthm, amssymb, graphicx}

\begin{document}

\title{Test of Particle-Assisted Tunneling for Strongly Interacting Fermions%
\\
in an Optical Superlattice}
\author{T. Goodman, L.-M. Duan}
\affiliation{FOCUS center and MCTP, Department of Physics, University of Michigan, Ann
Arbor, MI 48109}

\begin{abstract}
Fermions in an optical lattice near a wide Feshbach resonance are expected
to be described by an effective Hamiltonian of the general Hubbard model
with particle-assisted tunneling rates resulting from the strong atomic
interaction [Phys. Rev. Lett. 95, 243202 (2005)]. Here, we propose a scheme
to unambiguously test the predictions of this effective Hamiltonian through
manipulation of ultracold atoms in an inhomogeneous optical superlattice.
The structure of the low-energy Hilbert space as well as the particle
assisted tunneling rates can be inferred from measurements of the
time-of-flight images.
\end{abstract}

\maketitle

Fermions in an optical lattice near a wide Feshbach resonance provide one of
the most complicated systems for ultracold atoms \cite%
{ref1,ref2,ref3,ref4,duan1,ref6,ref5,ref6',duan2}. The strong atomic
interaction induced by Feshbach resonance populates many excited lattice
bands and causes direct couplings between neighboring sites. To understand
this important system, one needs to have a theoretical model. An effective
Hamiltonian has been proposed in Refs. \cite{duan1, duan2} based on the
arguments of the low-energy Hilbert space structure, which offers a
significant simplification for description of this system. The effective
Hamiltonian takes the form of the general Hubbard model (GHM), where the
effects of the multiband populations and the direct neighboring interactions
are incorporated through the particle-assisted tunneling rates. It is
important to test this model by comparing its predictions with experimental
observations. However, such a comparison is usually difficult because of the
lack of exact solutions to the GHM and complications in real experimental
configurations (such as the inhomogeneity due to the global trap).

In this paper, we propose an experimental scheme to quantitatively test the
predictions of this effective model by manipulating strongly interacting
atoms in an optical superlattice. The optical superlattice provides a
powerful tool, which has been used in recent experiments for demonstration
of the spin super-exchange interaction \cite{bloch1,NIST,bloch2}. In the
experimental configuration with an inhomogeneous optical superlattice,
through manipulation of the lattice barrier and the external magnetic field,
we show that one can reconstruct the two-site dynamics from the measured
time-of-flight images. The measured dynamics can then be compared with the
exact prediction from the general Hubbard model, offering an unambiguous
testbed for this complicated system. The proposed measurement also allows a
complete empirical determination of all the parameters in the effective GHM.

We consider two component fermions (denoted by spin $\sigma =\uparrow
,\downarrow $) in an optical lattice near a wide Feshbach resonance.
Although in general many lattice bands get populated due to the strong
atomic interaction, we note in Refs. \cite{duan1,duan2} that for this system the
low energy states at each site are still restricted to only four
possibilities: either a vacuum denoted by $\left\vert 0\right\rangle $, or a
single atom with spin-$\sigma $ denoted by $a_{\sigma }^{\dagger }\left\vert
0\right\rangle $, or a dressed molecule in the ground state $\left\vert
d\right\rangle $ which consists of superpositions of two-atom states
distributed over many bands. All the other states (such as the three-atom
states or the dressed molecule excited states) are well separated in energy,
and therefore not relevant for low-temperature physics. Based on this
low-energy Hilbert space structure and general symmetry arguments, it is
shown in \cite{duan2} that the effective Hamiltonian takes the form of the
GHM:
\begin{align}
H& =\sum_{i}\left[ \left( U/2\right) n_{i}\left( n_{i}-1\right) -\mu
_{i}n_{i}\right]  \label{Ham} \\
& +\sum_{\left\langle i,j\right\rangle ,\sigma }\left[ t_{a}+g_{1}\left( n_{i%
\overline{\sigma }}+n_{j\overline{\sigma }}\right) +g_{2}n_{i\overline{%
\sigma }}n_{j\overline{\sigma }}\right] a_{i\sigma }^{\dagger }a_{j\sigma
}+H.c.  \notag
\end{align}%
where $n_{i}\equiv \sum_{\sigma }a_{i\sigma }^{\dagger }a_{i\sigma },$ $n_{i%
\overline{\sigma }}\equiv a_{i\overline{\sigma }}^{\dagger }a_{i\overline{%
\sigma }}$ ($\overline{\sigma }=\downarrow ,\uparrow $ for $\sigma =\uparrow
,\downarrow $), $U$ characterizes the effective on-site interaction (defined
as the energy shift of $\left\vert d\right\rangle $ with respect to the
two-atom state on different sites), $\mu _{i}$ is the chemical potential (we
keep its dependence on the site $i$ for convenience of the following
discussion, where a global trap induces a site dependent energy shift), $%
t_{a}$ is the conventional single-atom tunneling rate, and $g_{1}$ and $%
g_{2} $ denote the additional tunneling assisted with spin-$\overline{%
\sigma }$ atoms (those two terms come from the multi-band populations in the
$\left\vert d\right\rangle $ state and the direct neighboring atomic
interaction in the lattice \cite{duan1}). In this derivation, we have
mathematically mapped $\left\vert d\right\rangle $ to the double occupation
state $a_{i\uparrow }^{\dagger }a_{i\downarrow }^{\dagger }\left\vert
0\right\rangle $ \cite{duan2} (though their physical compositions are
different).

To see whether the effective GHM gives a good description of the
low-temperature physics for this system, it is important to test the
predictions of the GHM through experiments. To have an unambiguous test, it
is better to design a configuration such that the GHM allows exact
solutions. The optical superlattice provides such an opportunity. To produce 
the optical superlattice, one simply adds a 3-dimensional lattice 
$V_{2}=V_{20}\sum_{\alpha =x,y,z}\sin ^{2}(\pi \alpha /2a)$ to a lattice 
$V_{1}=V_{10}\sin ^{2}(\pi z/a - \varphi )$ in one spatial direction (say $z$), 
where the periodicity $2a$ of $V_2$ is twice that of $V_1$\cite{bloch1,NIST,bloch2}. 
If $V_{10}$ is sufficiently large relative to $V_{20}$, the superposition of these 
two potentials produces a series of double wells along the $z$ direction. The 
dynamics in each double well are independent of the others provided the barrier 
between wells (controlled by $V_{20}$) is sufficiently large. Taking the relative 
phase $\varphi$ to be nonzero introduces an energy bias $\delta _{12}$ 
between the minima of each double well.

\begin{figure}[tbp]
\includegraphics[height=4cm,width=8cm]{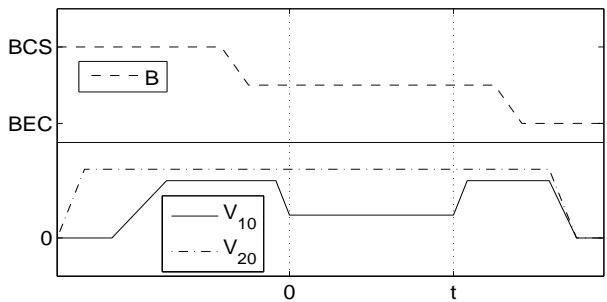}
\caption{The time sequences for the magnetic field ($B$) and the lattice
potentials ($V_{10}$ and $V_{20}$) to achieve state preparation, controlled
dynamics, and detection. (Note that the optical barriers are high during
sweeping of the $B$ field.)}
\label{potential}
\end{figure}

It is easy to calculate the dynamics in each double well from the GHM.
However, it is unclear how to directly measure the dynamics without
individual addressing of each well inside the lattice. The conventional
time-of-flight (ToF) images involve averages over all the potential wells.
These signals are further complicated by the presence of a global harmonic
trap $V_{g}=\sum_{\alpha =x,y,z}m\omega _{\alpha }^{2}\alpha ^{2}/2$
inevitable in an optical lattice, which makes each double well slightly
different. In the following, we show a scheme that can map out the detailed
dynamics in each double well from the measured ToF images even with the
presence of these complications.

The scheme here combines the control of both the optical potentials and the
magnetic field (see Fig. \ref{potential} for illustration). First, we need
to load each double well with a filling pattern that sets the initial
condition of the dynamics. This is achieved at the BCS limit of the
resonance. In this limit, the atoms are free fermions, and we can control
the filling pattern by choosing the total number $N=N_{\uparrow
}+N_{\downarrow }$ and the polarization $P=(N_{\uparrow }-N_{\downarrow })/N$%
. Then, we turn off all the inter-well dynamics by raising the barrier
(controlled by $V_{10}$ and $V_{20}$) and sweep the magnetic field to the
unitarity region. The sweeping speed $v$ is fast compared with the
inter-well coupling rate but small compared with the lattice gap of $V_{1}$
so that the levels in each single well adiabatically evolve. Near unitarity,
we turn on the inter-well dynamics for a duration $t$ by adjusting $V_{10}$ to 
lower the central barrier of each double well. These dynamics give information 
on the underlying strongly interacting Hamiltonian. To determine the final state 
after the dynamics, the central barrier is raised again, and the magnetic field 
is swept to the BEC limit with a speed similar to $v$. Depending on the particle 
number in each well, we have atoms or molecules or their mixture with negligible
interaction at the BEC limit. The ToF images for those atoms or molecules are then 
detected to determine the final state after the dynamics during time $t$.

\begin{figure}[tbp]
\includegraphics[height=3cm,width=8cm]{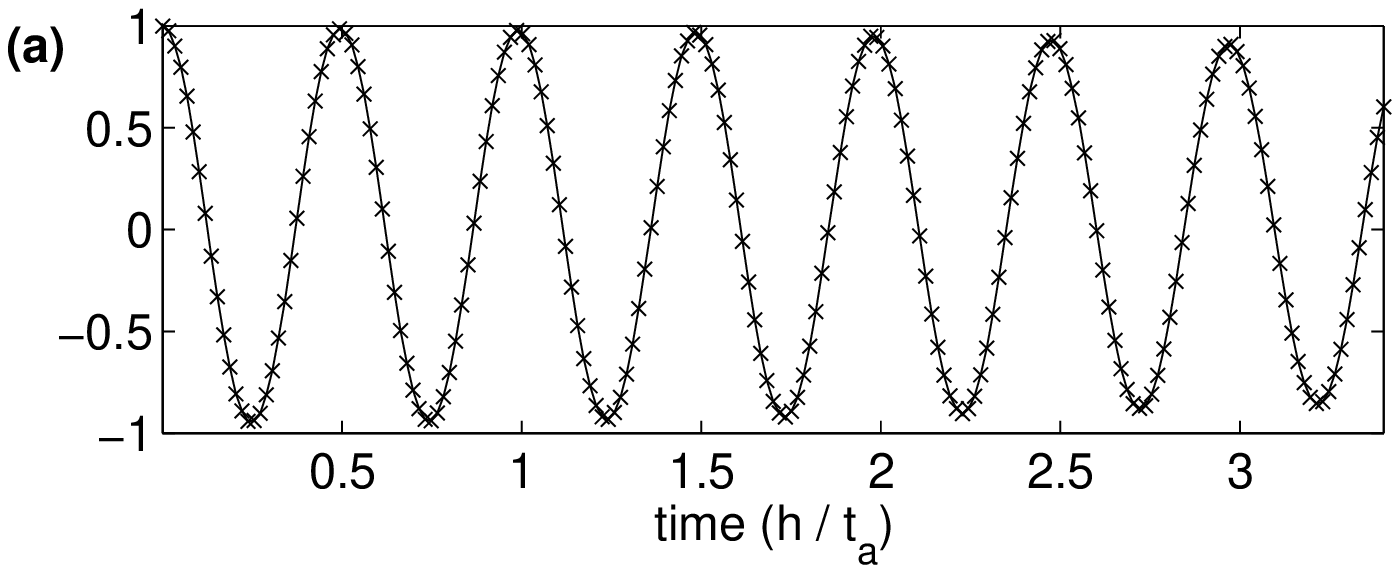} %
\includegraphics[height=4cm,width=8cm]{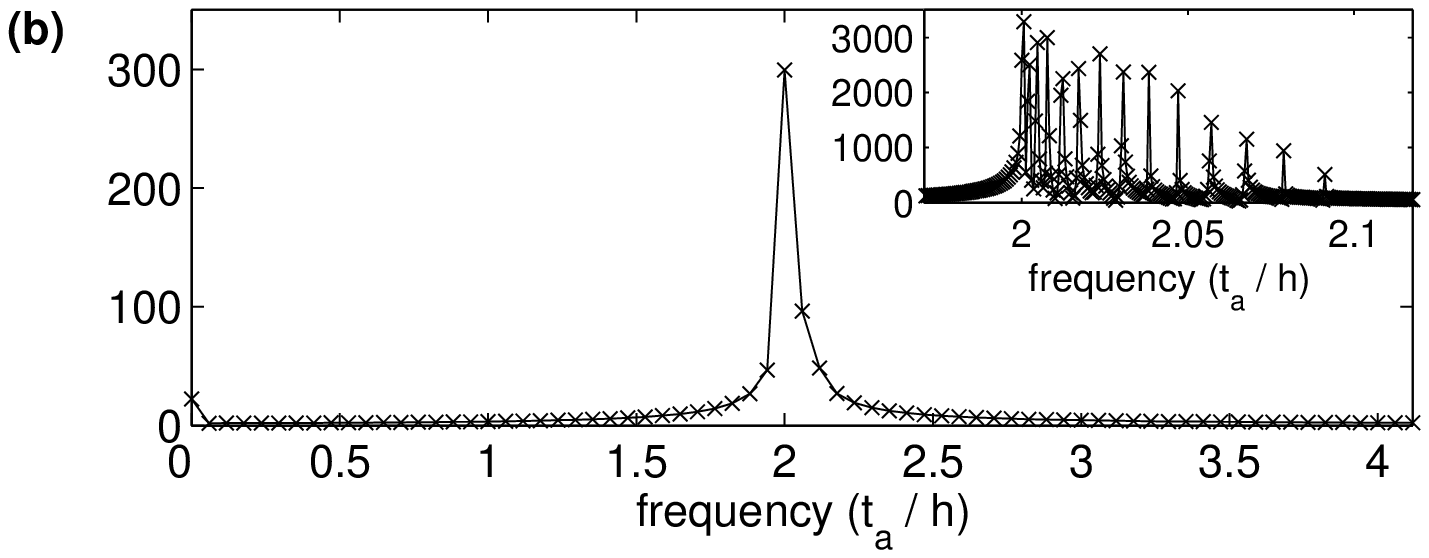}
\caption{Population difference between the left and right wells
$(N_L - N_R)/N$ for the case of one atom per double well.
{\bfseries (a):} Population difference vs. time (in the unit of
$h/t_a$). {\bfseries (b):} Fourier transform of the population
difference (frequency in the unit of $t_a/h$), calculated for a
time duration of $20h/t_a$ to give a frequency resolution of
$t_a/20h$. The peak occurs at a frequency $\protect\nu = 2 t_a / h $. {\bfseries %
Inset:} With the time duration increased to give a frequency
resolution of $t_a/2000h$, we see that there are actually many
peaks, corresponding to the
different frequencies $\Omega_{1i}$. $\Omega_{1i}$ depends on the $z$%
-coordinate and thus each peak corresponds to a different slice of
double wells parallel to the $z$-axis. The slices containing the
most occupied double wells are closest to $z = 0$. That is why
those peaks (which have the smallest $\Omega_{1i}$) dominate.
Because $t_a$ can be determined from the dominant peak, it is not
necessary to resolve the other smaller peaks. In calculation of
the inhomogeneity effect, we assume a spherical distribution with
a diameter of 30 occupied double wells, and take the following
typical values for the parameters: $t_a = h \times 170$ Hz, $m =
6.64 \times 10^{-26}$ kg (for $^{40}K$), $\protect\omega = 2
\protect\pi \times 80$ Hz, and $2a = 765$ nm.} \label{oscillation1}
\end{figure}

To test the prediction of particle-assisted tunneling, we need to compare
the free-atom hopping rate $t_{a}$ with $t_{a2}=t_{a}+g_{1}$ and $%
t_{a3}=t_{a}+2g_{1}+g_{2}$, where $t_{a2}$ and$\ t_{a3}$ correspond
respectively to the hopping rates of a spin-$\uparrow $ atom from the site $i
$ to $j$ when there is a spin-$\downarrow $ atom on one site or on both
sites. Let us first look at how to measure the free-atom hopping rate $t_{a}$
in the Hamiltonian (\ref{Ham}). For that purpose, we need one atom per
double well. By choosing the polarization $P=1$ and $V_{10}=0$ (so we have
at this stage single wells rather than double wells), the equilibrium
distribution of the free fermions at the BCS limit automatically gives this
configuration. The total atom number $N$ within the global harmonic trap $V_{g}$ 
needs to be below
\begin{equation}
N_{max}=\left( 4\pi /3\right) \left( E_{bg}/2m\omega^{2}a^{2}\right) ^{3/2}, \label{maxN}
\end{equation}
where $E_{bg}=2\sqrt{V_{10}\pi ^{2}\hbar ^{2}/8ma^{2}}$ is the band gap for
the lattice $V_{2}$ and we have assumed $\omega _{x}=\omega _{y}=\omega
_{z}\equiv \omega $. Then, one can adiabatically raise the potential $V_{10}$
with a bias $\delta _{12}$ so that the atom sits on the left-side well in
each double well. After raising $V_{10}$, $\delta_{12}$ is reduced to zero.
The system is then moved to the resonance region, and after turn-on of the 
dynamics for a duration $t$, the difference between the fraction of atoms in the 
left-side and the right-side wells over the whole harmonic trap is given by
\begin{equation}
\frac{N_{L}-N_{R}}{N}=1-\frac{2}{N}\sum_{i}\left( \frac{t_{a}}{\hbar \Omega
_{1i}}\right) ^{2}\sin ^{2}(\Omega _{1i}t),  \label{singleAtom}
\end{equation}%
where $\Omega _{1i}=\sqrt{\Delta _{i}^{2}+4t_{a}^{2}}/2\hbar ,$ $\Delta
_{i}\approx m\omega ^{2}az_{i}$. The summation of $i$ in Eq. (\ref{singleAtom})
is over all the occupied double wells in the global harmonic trap (with 
$z_{i}$ the $z$-coordinate of the center of the double well), and each double 
well has a slightly different bias $\Delta _{i}$ due to the trap potential $V_{g}$.
After the dynamics, in order to measure the populations $N_{L}$ and $N_{R}$,
the atom of the right-side well can be dumped into an excited vibrational
state (corresponding to the second band) of the left-side well by rapidly
raising the potential minimum of the right well relative to the left
(through control of the phase $\varphi $) \cite{bloch1,NIST,bloch2}. The
populations in different bands are then mapped out in the BEC limit through
measurement of the momentum distribution of free atoms with ToF imaging.
From the measured populations $N_{L}$ and $N_{R}$, one can easily determine
the tunneling rate $t_{a}$. Fig. \ref{oscillation1} (a) shows the typical
time evolution of $N_{L}-N_{R}$ from the dynamics, for which the oscillation
period is determined by $t_{a}$ and the damping is due to the inhomogeneity
from the global trap \cite{note}. In the frequency domain, the signal peaks
at $2t_{a}$, and the inhomogeneity causes many smaller peaks at frequencies
above that of the dominant peak (see Fig. \ref{oscillation1} (b)).

To measure the particle-assisted tunneling rate $t_{a2}$, we need two atoms
per double well, one spin-$\uparrow $ and one spin-$\downarrow $. This can
be achieved with the equilibrium distribution of free fermions at the BCS
limit by choosing $P\approx 0$, $V_{10}=0$, and the total atom number $N < 2
N_{max}$ (with $N_{max}$ defined in Eq. \ref{maxN}). The double well is
still turned on with a bias so that both of the atoms are prepared in the
left-side well. For the dynamics near resonance with the Hamiltonian (\ref%
{Ham}), the state at any time involves a superposition of three components:
a double occupation of the left or the right well, and a singlet state of
two atoms over the two wells. We can determine $t_{a2}$ as well as the
on-site interaction energy $U$ from the difference between the overall
fractions of double occupation of the left wells and of the right wells, $%
(N_{2L} - N_{2R})/(N_{2L} + N_{2R})$. (Here $N_{2L}$ and $N_{2R}$
are the total number of double wells in which the left and right
wells, respectively, are doubly occupied.) These fractions can be
directly measured at the BEC limit, where the double occupation of
a site is mapped to a molecule state, and the molecules in the
left and the right wells are distinguished through the band
mapping and the measurement of the momentum distribution (similar
to the discussion above for the atomic case). The single-atom
occupation of a well is mapped to an atomic state at the BEC
limit. Because of the large detuning between the atomic and the
molecular state, the atomic population does not contribute to the
time-of-flight imaging signal of the molecular fraction.

\begin{figure}[tbp]
\includegraphics[height=3cm,width=8cm]{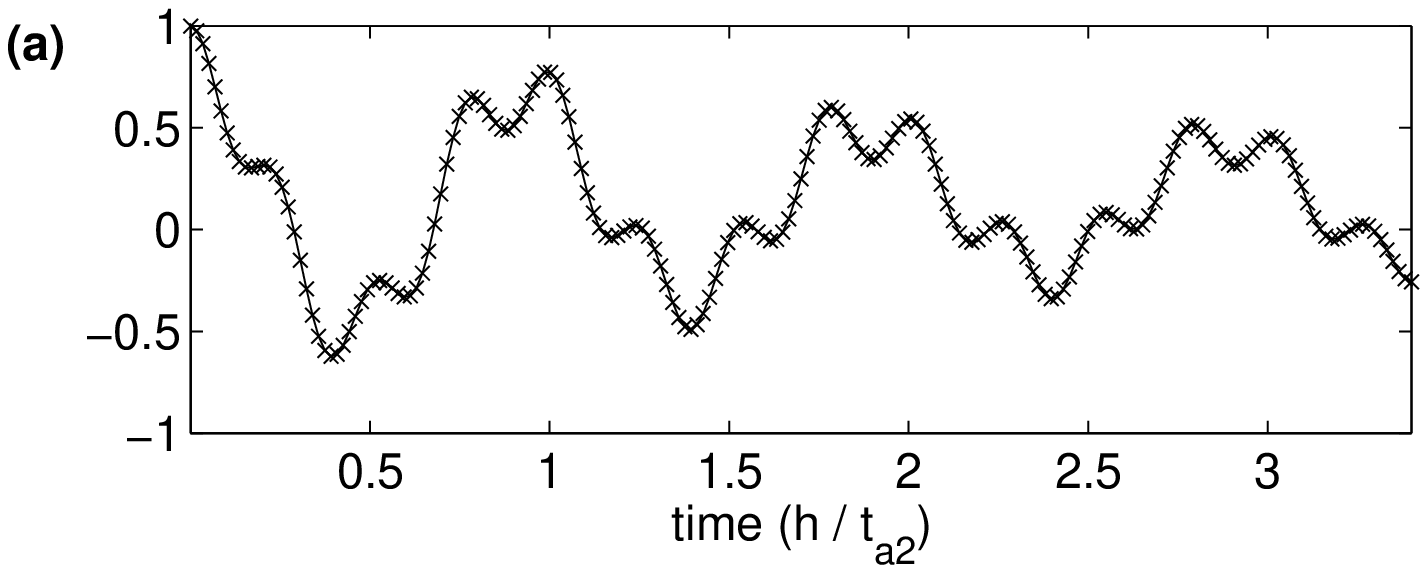} %
\includegraphics[height=3cm,width=8cm]{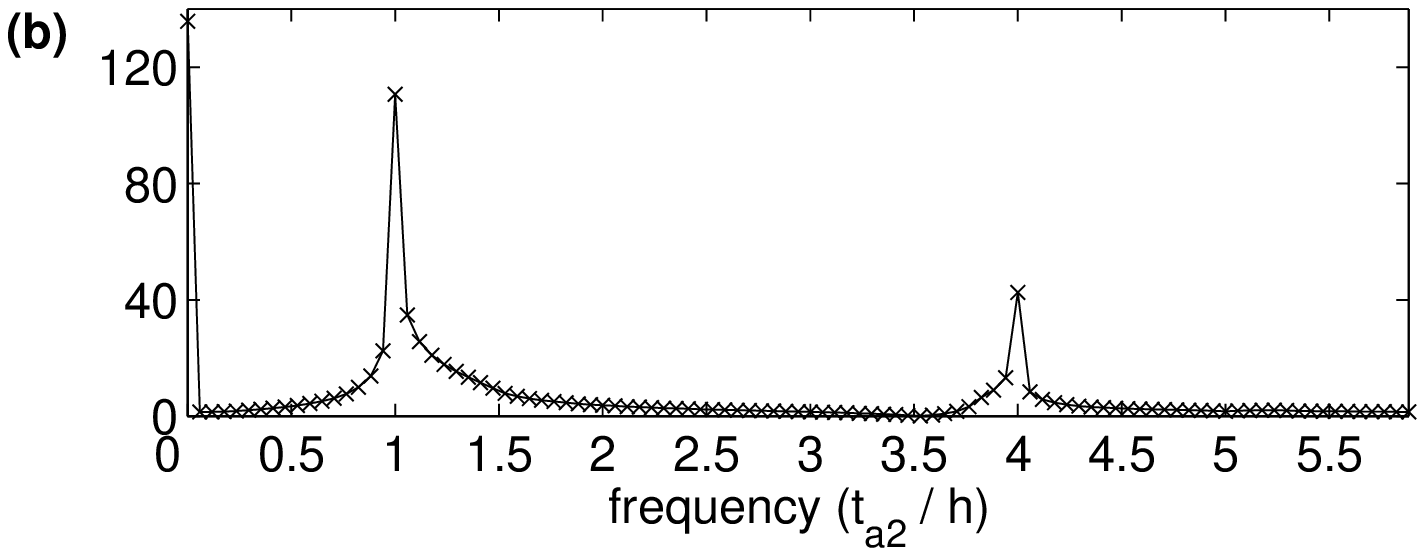}
\caption{The difference between the fractions of doubly occupied
left and right wells $(N_{2L}-N_{2R})/(N_{2L} + N_{2R})$ for the
case of two atoms per double
well. {\bfseries%
(a):} Population difference vs. time (in the unit of $h/t_{a2}$).
{\bfseries(b):} Fourier transform of the population difference
(the frequency resolution is $1/20$ in the unit of $t_{a2}/h$. The
peaks occur at a
frequencies $\protect\nu _{1}=\left( \protect\sqrt{U^{2}+16t_{a2}^{2}}%
-U\right) /2h$ and $\protect\nu _{2}=\left( \protect\sqrt{%
U^{2}+16t_{a2}^{2}}+U\right) /2h$ (in the figure we take
$U=3t_{a2}$ as an example). Increasing the frequency resolution
would reveal a series of smaller peaks on the high frequency side
of the large peaks (as in Fig. \protect\ref{oscillation1}), but it
is not necessary to resolve these smaller peaks to determine
$t_{a2}$ and $U$. } \label{oscillation2}
\end{figure}

The typical time evolution of $(N_{2L}-N_{2R})/(N_{2L} + N_{2R})$ is shown in Fig. \ref%
{oscillation2} (a). In the frequency domain (see Fig.
\ref{oscillation2} (b)), one can see two distinct primary peaks in
the Fourier transform, centered at $\left(
\sqrt{U^{2}+16t_{a2}^{2}}\pm U\right) /2$. The smaller peaks from
the inhomogeneity of the global harmonic trap do not obscure these
two dominant peaks. The frequencies at which these two peaks occur
can be understood by the fact that while the oscillation frequency
varies from well to well due to the $z$-dependent bias, there are
a greater number of occupied double wells near $z=0$ than for any
other $z$-coordinate. Thus, the dominant peaks correspond to the
zero bias case, where we have
\begin{equation}
\frac{N_{2L}^{(0)}-N_{2R}^{(0)}}{N_{2L}^{(0)}+N_{2R}^{(0)}}=\frac{\Omega _{+}%
}{\Omega }\cos \left( \Omega _{-}t\right) +\frac{\Omega _{-}}{\Omega }\cos
\left( \Omega _{+}t\right) ,  \label{doubleAtom}
\end{equation}%
with $\Omega _{\pm }=\left( \hbar \Omega _{2i}\pm U\right) /2\hbar $ and $%
\Omega _{2i}=\sqrt{U^{2}+16t_{a2}^{2}}/\hbar $.

\begin{figure}[tbp]
\includegraphics[height=4cm,width=8cm]{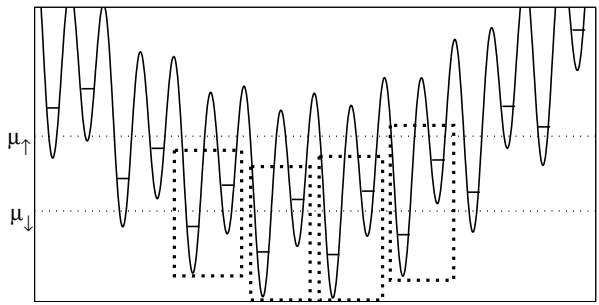}
\caption{The superlattice configuration to achieve three atoms per double
well. This is obtained by turning on the lattice potentials $V_1$ and $V_2$
simultaneously with relative phase $\protect\varphi > 0$, producing double
wells with a non-zero potential bias between the left and right wells. (The
overall harmonic potential is exaggerated for illustration purposes). In the
figure, the solid line in each well corresponds to the lowest level, the
long dotted lines correspond to the Fermi surfaces for $\uparrow$-atoms and $%
\downarrow$-atoms (with Fermi energies $\protect\mu_\uparrow$ and $\protect%
\mu_\downarrow$), which differ due to the polarization $P > 0$, and the dotted
rectangles indicate those double wells that are occupied by two $\uparrow$-atoms 
and one $\downarrow$-atom. The $\protect\mu_\uparrow$ and $\protect\mu_\downarrow$
are chosen such that $\downarrow$-atoms only occupy the left wells while $%
\uparrow$-atoms occupy both wells. This is the initial configuration needed
to measure the hole hopping rate $t_{a3}$. There is also the possibility of
additional $\uparrow$-atoms further from the center of the trap, but the
measured molecule signal is only sensitive to double wells containing \emph{%
both} $\uparrow$-atoms and $\downarrow$-atoms. With the conditions given in
the text, we insure that the only such double wells are those with two $%
\uparrow$-atoms and one $\downarrow$-atom.}
\label{threeAtoms}
\end{figure}

One can also measure the parameter $g_{2}$ in the Hamiltonian (\ref{Ham}),
which requires three atoms per double wells (two spin-$\uparrow $, one spin-$%
\downarrow $). One can consider this case as a single spin-$\uparrow$ hole
in each double wells, with a hole hopping rate of $t_{a3}$. This hopping rate
can be measured by the same method as for measurement of the free atom
hopping rate $t_{a1}$. To prepare three atoms per double wells, one can
consider the free fermion distribution at the BCS limit in an asymmetric
double-well lattice with a bias $\delta _{12}$ controlled by the phase shift
$\varphi$. We would like to have two atoms (one spin-$\uparrow $ and one
spin-$\downarrow $) in the deep wells and one spin-$\uparrow $ atom in the
shallow wells as shown in Fig. \ref{threeAtoms}. This can be achieved by
choosing the polarization $P$ and bias $\delta_{12}$ so that the atom
numbers satisfy $N_{\uparrow} > \left( 2\sqrt{2} + 1 \right) N_0$ and $%
N_{\downarrow} < N_0$, where $N_0 = \left( 4\pi /3\right) \left(\delta_{12}/2m\omega^2 a^2\right)^{3/2}$. 
These relations were derived by requiring that $N_\uparrow$ be great enough 
that every double well which contains a $\downarrow$-atom must also contain at 
least two $\uparrow$-atoms (so that the molecule signal corresponds only to 
double wells containing three atoms), and also requiring that no double well 
contain more than one $\downarrow$-atom. (Note that it is not sufficient to 
require a polarization $P \geq 1/3$, since this could be achieved with an 
inner core of double wells containing one $\uparrow$-atom and one $\downarrow$-atom, 
surrounded by a shell of double wells containing only an $\uparrow$-atom.) $N_\uparrow$ 
must also be small enough that there are no more than two $\uparrow$ atoms per
double well in the center of the trap. This condition can be met along with
the above conditions provided the band gap of the lattice is sufficiently
great.

A key assumption in deriving the Hamiltonian (\ref{Ham}) is that in the
strongly interacting region there is a significant energy gap (of the order
of the band gap) which separates the four low energy states on each site
from the other higher energy states \cite{duan1,duan2}. With the
superlattice technique, one can directly test this assumption and measure
the energy gap. Given this energy gap, if we fill each site with two atoms,
there will be no dynamics as long as the atomic tunneling rate between the
two sites is small compared with the band gap energy. To fill each site with
two atoms (one spin-$\uparrow $, one spin-$\downarrow $), we can start with
the free-fermion distribution in the BCS limit, choosing the polarization $%
P\approx 0$ and total atom number $N < 4 N_{max}$ (with $N_{max}$
as defined in Eq. \ref{maxN}). We then adiabatically turn on $V_1$
and $V_2$ simultaneously while keeping a constant ratio
$V_{10}/V_{20} > 1$. With this filling pattern, we should see no
dynamics in the strongly interacting region, so the atomic
distribution over the two sites (which will be mapped to the
molecular population distribution in the BEC limit) will not
change with the evolution time $t$. One can also tilt the
double-well lattice by tuning the bias $\delta _{12}$, and measure
what is the critical $\delta _{12}$ to turn on the two-site
dynamics in the population distribution. The measured critical
$\delta _{12}$ will give an estimate of the energy gap to excite
the system to the high energy states.

In summary, we have described a scheme to test in a controllable
fashion the predictions of an effective Hamiltonian for strongly
interacting fermions in an optical lattice. With the superlattice
technique, one can directly test the key assumption in derivation
of the Hamiltonian, and can measure the physical parameters to
confirm the particle-assisted tunneling. This scheme provides a
quantitative testbed to compare theory with experiments in the
strongly interacting region.

This work is supported under the MURI program and under ARO Award
W911NF0710576 with funds from the DARPA OLE Program.


\begin{thebibliography}{99}
\bibitem{note} Note that in this case, the dynamics should be the same as
the bosonic case, and the latter has been measured in recent experiments
\cite{bloch1}.
\end{thebibliography}
\end{document}